\documentclass[12pt,epsf]{article}
\usepackage{graphicx}
\usepackage{epsfig}
\usepackage{amsmath}
\usepackage{subfigure}
\usepackage{slashbox}
\begin{document}

\def\a{\alpha}
\def\b{\beta}
\def\c{\varepsilon}
\def\d{\delta}
\def\e{\epsilon}
\def\f{\phi}
\def\g{\gamma}
\def\h{\theta}
\def\k{\kappa}
\def\l{\lambda}
\def\m{\mu}
\def\n{\nu}
\def\p{\psi}
\def\q{\partial}
\def\r{\rho}
\def\s{\sigma}
\def\t{\tau}
\def\u{\upsilon}
\def\v{\varphi}
\def\w{\omega}
\def\x{\xi}
\def\y{\eta}
\def\z{\zeta}
\def\D{\Delta}
\def\G{\Gamma}
\def\H{\Theta}
\def\L{\Lambda}
\def\F{\Phi}
\def\P{\Psi}
\def\S{\Sigma}

\def\o{\over}
\def\beq{\begin{eqnarray}}
\def\eeq{\end{eqnarray}}
\newcommand{\lsim}{\raisebox{0.6mm}{$\, <$} \hspace{-3.0mm}\raisebox{-1.5mm}{\em $\sim \,$}}
\newcommand{\gsim}{\raisebox{0.6mm}{$\, >$} \hspace{-3.0mm}\raisebox{-1.5mm}{\em $\sim \,$}}

\newcommand{\vev}[1]{ \left\langle {#1} \right\rangle }
\newcommand{\bra}[1]{ \langle {#1} | }
\newcommand{\ket}[1]{ | {#1} \rangle }
\newcommand{\EV}{ {\rm eV} }
\newcommand{\MeV}{ {\rm MeV} }
\newcommand{\GeV}{ {\rm GeV} }
\newcommand{\TeV}{ {\rm TeV} }
\def\diag{\mathop{\rm diag}\nolimits}
\def\Spin{\mathop{\rm Spin}}
\def\SO{\mathop{\rm SO}}
\def\O{\mathop{\rm O}}
\def\SU{\mathop{\rm SU}}
\def\U{\mathop{\rm U}}
\def\Sp{\mathop{\rm Sp}}
\def\SL{\mathop{\rm SL}}
\def\tr{\mathop{\rm tr}}

\def\IJMP{Int.~J.~Mod.~Phys. }
\def\MPL{Mod.~Phys.~Lett. }
\def\NP{Nucl.~Phys. }
\def\PL{Phys.~Lett. }
\def\PR{Phys.~Rev. }
\def\PRL{Phys.~Rev.~Lett. }
\def\PTP{Prog.~Theor.~Phys. }
\def\ZP{Z.~Phys. }

\def\Z{\mathcal{Z}}
\def\W{\Omega}

\baselineskip 0.7cm

\begin{titlepage}

\begin{flushright}
RESCEU-20/10\\
IPMU10-0154\\
UT-10-16
\end{flushright}

\vskip 1.35cm
\begin{center}
{\large \bf
Primordial Black Hole as a Source of the Boost Factor
}
\vskip 1.2cm
Ryo Saito$^{1, 2}$ and Satoshi Shirai$^{1,3}$
\vskip 0.4cm

{\it
$^1$Department of Physics, University of Tokyo, 
Tokyo 113-0033,
Japan\\
$^2$
RESCEU,
University of Tokyo,Tokyo 113-0033, Japan,

$^3$IPMU,
University of Tokyo, 
Chiba 277-8586, 
Japan\\
}

\vskip 1.5cm

\abstract{
Primordial black holes (PBHs) accumulate weakly interacting massive particles (WIMPs) around them and form ultracompact minihalos (UCMHs), if the WIMP is a dominant component of the dark matter (DM). In this paper, we discuss that the UCMHs seeded by the PBHs with sub-earth mass enhance the WIMP annihilation in the present Universe and can successfully explain the positron and/or electron excess in cosmic ray observed by PAMELA/Fermi experiments. The signal is very similar to that from a decaying dark matter, which can explain the PAMELA and/or Fermi anomaly without conflict with any constraints as long as the decay mode is proper. In this scenario, the boost factor can be as large as $10^5$. In addition, we discuss testability of our scenario by gamma-ray point source and gravitational-wave experiments.
}
\end{center}
\end{titlepage}

\setcounter{page}{2}

\section{Introduction}
The origin of the dark matter (DM) is one of the most challenging problems in both particle physics and cosmology.
In the context of beyond the Standard Model (BSM), a weakly interacting massive particle (WIMP) is a good candidate of the DM.
In many models, DM is considered as thermal relic, whose abundance is determined by its annihilation cross section:
\beq
\Omega_{\rm{WIMP}} h^2 \simeq 0.1 \left(\frac{\langle \sigma v \rangle}{3\times10^{-26}~{\rm cm^3 s^{-1}}}\right)^{-1}, \label{eq:cross}
\eeq
where $\langle \sigma v \rangle$ is thermal average of a product of the DM annihilation cross section and velocity.
 To explain the present DM abundance, $\Omega_{\rm DM}h^2 \simeq 0.1$ \cite{Komatsu:2010fb}, the annihilation cross section is required to be $3\times10^{-26}~{\rm cm^3 s^{-1}}$,
which is a reasonable value for $m_{\rm WIMP}={\cal O}(100)$ GeV in many BSMs.

The annihilation processes also occur in the galactic halo in the present Universe and 
can be a source of the cosmic ray.
Recently, 
the PAMELA collaboration has reported the excess of the cosmic ray positron \cite{Adriani:2008zr}.
If the origin of the anomaly is the DM annihilation, the required annihilation cross section is very large, typically 
$\langle \sigma v \rangle \sim 10^{-24}~ {\rm cm^3 s^{-1}}$, which is ${\cal O}(100-1000)$ times larger than the cross section
that accounts for the present DM abundance as seen in Eq.~(\ref{eq:cross}).
Therefore,  one needs to enhance the annihilation processes to explain the PAMELA anomaly.
This enhancement is parametrized by so-called ``boost factor.''

There are two kinds of the boost factor.
One is originated from particle physics.
For example, the large cross section can be achieved through the Sommerfeld enhancement \cite{ArkaniHamed:2008qn} or the Breight-Wigner enhancement \cite{Ibe:2008ye}.
In such cases, the annihilation cross section has large velocity dependence and the large cross section is realized 
at lower velocities.
Another possibility in particle physics is non-thermal production of the DM.
In this case, the DM is produced by non-thermal processes such as decay of heavy particles.
Therefore, the annihilation cross section does not determine the DM abundance as in  Eq.~(\ref{eq:cross}) and can have a large value.
However, there are severe constraints in both cases
since a large amount of energy injected by the DM annihilation with such a large cross section can change the history of the Universe, such as the Big-Bang Nucleosynthesis (BBN) \cite{Hisano:2009rc} or
Cosmic Microwave Background (CMB) \cite{Kanzaki:2009hf}. In addition,  gamma ray observations give strong constraints on such a DM \cite{Bertone:2008xr, Abdo:2010dk}.

The other source of the boost factor is originated  from astrophysics.
The annihilation rate is proportional to $n^2_{\rm DM}$, where $n_{\rm DM}$ is number density of the DM.
Therefore some clumps, in which the DM density is higher than one of the galactic halo, if any, can be a source of the boost factor.
A candidate of such clumps is DM subhalos, which are predicted to be contained in the galactic halo by numerical simulations \cite{Diemand:2008in, Springel:2008cc}. A large subhalo, however, has to lie near the Solar System to explain the PAMELA anomaly and such cases are improbable \cite{Brun:2009aj}
.

In this paper, we propose a new source of the boost factor, which is originated from minisize halos seeded by primordial black holes (PBHs) \cite{Hawking:1971ei, Carr:1974nx}, called ultracompact minihalos (UCMHs) \cite{Mack:2006gz, Ricotti:2007jk, Ricotti:2009bs}. The PBHs can be formed in the very early Universe if there exist density fluctuations of order of unity. If they make up only a small amount of the DM at the formation time, they subsequently accrete the surrounding WIMP DM and form minihalos, UCMHs, around them. The WIMP density in the UCMHs is so high that a small fraction of them can lead to significant enhancement of the cosmic ray signal \cite{Lacki:2010zf, Scott:2009tu}. The boost factor is determined by the UCMH abundance and can be as large as ${\cal O}(10^5)$ if there are the UCMHs whose abundance is comparable to that of the DM in the present Universe. In this scenario, total flux of the cosmic ray is given by integration of each UCMH's flux. Hence the signal is very similar to that from a decaying DM, which has no conflict with the BBN, CMB \cite{Zhang:2007zzh} and the gamma ray observations \cite{Nardi:2008ix, Chen:2009uq}.

This paper is organized as follows. In Section \ref{sec:enhance} we describe the annihilation enhancement in the UCMHs and estimate the annihilation rate per UCMH. In Section \ref{sec:boost} we estimate the boost factor induced by the UCMHs and give the UCMH abundance necessary to explain the PAMELA anomaly. Section \ref{sec:point} discusses prospects and constraints from gamma-ray point source observations. Our summary and conclusion are presented in Section \ref{sec:summary}. 

\section{Annihilation Enhancement in UCMHs}\label{sec:enhance}
First, we introduce a WIMP density profile in the UCMH seeded by the PBH and estimate the annihilation rate per UCMH, $\Gamma$, based on the discussion in Ref.~\cite{Lacki:2010zf}.

\subsection{WIMP Density Profile in the UCMH}

The PBHs formed in the radiation-dominated era have masses of order of the horizon mass at their formation time:
\beq
	M_{\rm PBH} \simeq M_{\oplus}\left(\frac{T}{100~{\rm GeV}}\right)^{-2},
\eeq
where $T$ is temperature of the Universe\footnote{Here, $M_{\oplus}$ is the earth mass, $6 \times 10^{27}~{\rm g}$. We also denote the solar mass, $2 \times 10^{33}~{\rm g}$, as $M_{\odot}$}. The PBHs lighter than $10^{15}~{\rm g}$ have evaporated away by now due to the Hawking radiation, hence, they have to be heavier than $10^{15}~{\rm g}$ to remain in the present Universe. The PBH accumulates the WIMP DM on the order of its original mass until the matter-radiation equality, $z=z_{\rm eq}$, and the accreted mass increases as $\propto (1+z)^{-1}$ during the matter-dominated era \cite{Mack:2006gz, Ricotti:2007jk, Bertschinger:1985pd}:
	\beq
	M_{\rm h}(z)=M_{\rm PBH}\left( \frac{1+z}{1+z_{\rm eq}} \right)^{-1}.
	\eeq
After the structure formation, $z \simeq 30$, the growth of the UCMHs proceeds depending on the environment around them. The accretion can also stop at $1+z_{\emptyset} \equiv f_{{\rm PBH},i}(1+z_{\rm eq})/(1-f_{{\rm PBH},i})$, when almost all of the WIMPs are accreted. Here, $f_{{\rm PBH},i}$ is the initial fraction of the PBHs in the DM energy density, which should be much less than unity to accrete the WIMPs initially. In this case, the UCMHs constitute the present DM. Note that the DM annihilation cross section should be of order of the thermal one even in this case because the UCMH mass is dominated by the WIMPs.

The halo radius at the redshift $z$, where the WIMP density profile truncates, is given by \cite{Mack:2006gz, Ricotti:2007jk, Bertschinger:1985pd}
\begin{equation}\label{eq:rtr}
R_{\rm tr}(z) = 18 ~{\rm AU}\left(\frac{M_{\rm h}(z)}{M_{\oplus}}\right)^{1/3}\left(\frac{1+z}{1+z_{\rm eq}}\right)^{-1}.
\end{equation}
 We parametrize the WIMP energy density profile in the UCMH as,
\beq\label{eq:profile}
	\rho_{\rm WIMP}(r) \propto
	\begin{cases}
		r^{-\nu_i} & (r \lsim R_{\rm eq}), \\
		r^{-\nu_o} & (r \gsim R_{\rm eq}),
	\end{cases}
\eeq
where we have denoted $R_{\rm tr}(z_{\rm eq})$ as $R_{\rm eq}$, above which the accreted mass exceeds the PBH mass. Here, the indices are assumed to satisfy $0 < \nu_i \lsim 1.5$ and $\nu_o > 1.5$. In the inner part of the halo, a large part of the halo is made up of the WIMPs accreted at high redshift because the WIMPs accreted later pass through there so quickly. Therefore, the orbits of the WIMPs in the inner region, $r \lsim R_{\rm eq}$, where the PBH dominates the mass, are insensitive to the density profile there. If the angular momentum of the WIMPs is sufficiently small, they free-fall into the PBH and the density profile scales as $\rho_{\rm WIMP} \propto r^{-1.5}$. On the other hand, in the outer part, $r \gsim R_{\rm eq}$, where the halo mass exceeds the PBH mass, the WIMPs feel force from the mass accreted at earlier time. In this region, the profile is expected to scale as the self-similar one, $\rho_{\rm WIMP} \propto r^{-2.25}$ \cite{Bertschinger:1985pd}. Hence, our assumption is satisfied in this case. The dark matter density at $R_{\rm eq}$ is determined irrespective of the values of the PBH mass as $\rho_{\rm eq} \equiv 1.8 \times 10^7~ {\rm GeV/cm^3}$, which is ${\cal O}(10^{7-8})$ larger than the local DM density, $\rho_{\odot} \simeq 0.3~{\rm GeV/cm^3}$.

\subsection{Annihilation Rate}

The WIMP annihilation rate per UCMH, $\Gamma$, is given by
\beq
	\Gamma \equiv \frac{1}{m_{\rm WIMP}^2}\int_{R_{\rm min}}^{R_{\rm tr}(z_{\rm s})}2\pi r^2 \rho_{\rm WIMP}^2(r)\langle \sigma v \rangle {\rm d}r,
\eeq
where $z_{\rm s}$ denotes the redshift when the UCMHs stop growing, which we assume to be $\max(30,z_{\emptyset})$, and $R_{\rm min}$ is a minimum-radius cutoff. For the WIMP density profile (\ref{eq:profile}), $\Gamma$ is determined by the DM annihilation rate in the vicinity of $r \simeq R_{\rm eq}$. Therefore, it is convenient to rewrite $\Gamma$ as,
\begin{equation}
\Gamma = 1 \times 10^{29}~{\rm s}^{-1}
I_{\rm profile}\left(\frac{m_{\rm WIMP}}{100~{\rm GeV}}\right)^{-2}
\left(\frac{M_{\rm PBH}}{M_{\oplus}}\right)
\left(\frac{\left<\sigma v\right>}{3\times 10^{-26}~{\rm cm^3s^{-1}}}\right),
\end{equation}
where $I_{\rm profile}$ is defined as
\beq
I_{\rm profile} \equiv \int_{R_{\rm min}/R_{\rm eq}}^{R_{\rm tr}(z_{\rm s})/R_{\rm eq}} \hat{r}^2 \left(\frac{\rho_{\rm WIMP}}{\rho_{\rm eq}}\right)^2{\rm d}\hat{r} \qquad (\hat{r} \equiv r/R_{\rm eq}).
\eeq
 The profile-dependent integral, $I_{\rm profile}$, can be determined irrespective of $R_{\rm tr}(z_{\rm s})$ and $R_{\rm min}$ except in the case $\nu_i \simeq 1.5$ as
\beq
I_{\rm profile} \simeq \frac{\nu_o-\nu_i}{(\nu_o-1.5)(1.5-\nu_i)}.
\eeq
 In the case $\nu_i \simeq 1.5$, $I_{\rm profile}$ depends on $R_{\rm min}$ only logarithmically;
\beq
	I_{\rm profile} \simeq \ln\left(\frac{R_{\rm eq}}{R_{\rm min}}\right).
\eeq

Here, the minimum-radius cutoff, $R_{\rm min}$, can be estimated by considering a flattening of the density profile due to the WIMP annihilation; the annihilation rate is high enough to reduce the density in the inner high-dense region \cite{Vasiliev:2007vh}. The shallower profile develops from a radius where the density becomes comparable to
\beq
	\rho_{\rm a} = \frac{m_{\rm WIMP}}{\langle \sigma v\rangle(t_0-t_i)} = 8 \times 10^9 ~{\rm GeV/cm^3}\left(\frac{m_{\rm WIMP}}{100~{\rm GeV}}\right)\left(\frac{\langle \sigma v\rangle}{3 \times 10^{-26}~{\rm cm^3/s}} \right)^{-1},
\eeq
where $t_0-t_i$ is time elapsed from the formation of the UCMH, $4 \times 10^{17}~{\rm s}$. Therefore, the minimal-radius cutoff, $R_{\rm min}$, can be estimated to be
\beq \label{eq:rmin}
\frac{R_{\rm min}}{R_{\rm eq}} = \left(\frac{\rho_{\rm a}}{\rho_{\rm eq}}\right)^{-\frac{2}{3}} = 0.017\left(\frac{m_{\rm WIMP}}{100~{\rm GeV}}\right)^{-\frac{2}{3}}\left(\frac{\langle \sigma v\rangle}{3 \times 10^{-26}~{\rm cm^3/s}} \right)^{\frac{2}{3}},
\eeq
which corresponds to $I_{\rm profile} = {\cal O}(1)$.

We should also consider propagation of the electron/positron in the UCMH; they lose their energy by scattering the baryons. For this purpose, we compare the radiation length of the high energy electron in hydrogen, $X_0 \simeq 60~{\rm g/cm^2}$ \cite{Amsler:2008zzb}, with the corresponding length of the baryons within the UCMH,
\beq
f_B\int_{R_{\rm min}}^{R_{\rm tr}(z_{\rm s})} \rho_{\rm WIMP} {\rm d}r \simeq 2f_B\rho_{\rm eq}R_{\rm eq}\left(\frac{R_{\rm min}}{R_{\rm eq}}\right)^{-1/2},
\eeq
for $\nu_i \simeq 1.5$. Here, we have assumed that the baryons distribute as $\rho_{B} = f_B\rho_{\rm WIMP}$ in the vicinity of $r \simeq R_{\rm eq}$. Substituting Eq.(\ref{eq:rmin}), we obtain
\beq
f_B\int_{R_{\rm min}}^{R_{\rm tr}(z_{\rm s})} \rho_{\rm WIMP} {\rm d}r \simeq 0.13~{\rm g/cm^2}f_B\left(\frac{M_{\rm PBH}}{M_{\oplus}}\right)^{\frac{1}{3}}\left(\frac{m_{\rm WIMP}}{100~{\rm GeV}}\right)^{\frac{1}{3}}\left(\frac{\langle \sigma v\rangle}{3 \times 10^{-26}~{\rm cm^3/s}} \right)^{-\frac{1}{3}},
\eeq
which is smaller than the radiation length, $X_0$. Hence, we can neglect the energy loss of the electron/positron in the UCMH.

In conclusion, the order of magnitude of $\Gamma$ is,
\beq
	\Gamma = {\cal O}(10^{29})~{\rm s}^{-1}\left(\frac{m_{\rm WIMP}}{100~{\rm GeV}}\right)^{-2}
\left(\frac{M_{\rm PBH}}{M_{\oplus}}\right)
\left(\frac{\left<\sigma v\right>}{3\times 10^{-26}~{\rm cm^3s^{-1}}}\right).
\eeq

In the next section, we estimate the abundance of the UCMHs required to explain the PAMELA anomaly by using this rate.

\vspace*{.5cm}

Before ending this section, we have to comment on feasibility of our setup. Though we have considered only the UCMHs seeded by the PBHs as sources of the cosmic ray, a number of the UCMHs could be formed without producing PBHs \cite{Ricotti:2009bs}. Density fluctuations with amplitude $\gsim {\cal O}(10^{-3})$ grow in the matter-dominated era and can collapse before $z \sim 10^3$. These fluctuations are too small to produce the PBHs, but can seed a much larger number of the UCMHs than the PBHs do. Because these UCMHs have no core, the self-similar density profile, $\rho_{\rm WIMP} \propto r^{-2.25}$, is expected to extend to the inner region.  In this case, the annihilation rate is estimated to be $\Gamma \sim \rho_{\rm a}^2R_{\rm min}^3$. The minimum-radius cutoff scales as $R_{\rm min} \propto \rho_{\rm a}^{-\frac{4}{9}}$, so that the annihilation rate strongly varies in time as $\Gamma \propto (1+z)$. Then, the rate is severely constrained by the recent Fermi observation of the extra galactic gamma ray and would moreover violate the constraints from the CMB observations \cite{Zhang:2007zzh}. 

This problem can be avoided if the PBH mass is so small that the accompanying small-scale fluctuations which would grow to form the coreless UCMHs eventually are wiped out due to collisional damping and free streaming of the WIMPs \cite{Hofmann:2001bi}. Though the collisional damping and the free streaming have comparable length scales, the latter is larger by a factor of ${\cal O}(10)$. The comoving length scale of the free streaming is given by
\beq
	l_{\rm fs} \simeq 1~{\rm pc}\left(\frac{m_{\rm WIMP}}{100~{\rm GeV}}\right)^{-\frac{1}{2}}\left(\frac{T_{\rm kd}}{10~{\rm MeV}}\right)^{-\frac{1}{2}},
\eeq
where $T_{\rm kd}$ is the temperature at kinetic decoupling of the WIMPs. The value of $T_{\rm kd}$ depends on the model, but is typically $T_{\rm kd} \sim 10~{\rm MeV}$. On the other hand, a PBH with mass $M_{\rm PBH}$ is formed by the gravitational collapse of the density fluctuations with a comoving scale,
\beq
	l_{\rm PBH} \simeq 2 \times 10^{-3}~{\rm pc}\left(\frac{M_{\rm PBH}}{M_{\oplus}}\right)^{\frac{1}{2}}.
\eeq
 Therefore, if the PBH mass is sufficiently small,
\beq
	M_{\rm PBH} < 2\times 10^6 M_{\oplus}\left(\frac{m_{\rm WIMP}}{100~{\rm GeV}}\right)^{-1}\left(\frac{T_{\rm kd}}{10~{\rm MeV}}\right)^{-1},
\eeq
possible constraints from the CMB observations can be evaded.

We should also check whether the angular momentum of the WIMPs can be neglected because the equations in this section have been derived in the approximation of quasi-radial infall. The WIMPs have velocity dispersions induced gravitationally and thermally. The former is given by,
\beq \label{eq:gravitational}
	\sigma_{\rm g}^2 \equiv \frac{{\cal H}^2\Omega_{\rm m}^{1.1}}{2\pi^2}\int_{\gsim l^{-1}}P(k){\rm d}k,
\eeq
at comoving scale $l$ where ${\cal H}$ is the comoving Hubble parameter and $P(k)$ is the matter power spectrum \cite{Peebles:1980}. Hence, it is determined by the matter power spectrum on scales smaller than $l$. The WIMPs turned around at comoving radius,
\beq
 l_{\rm ta} \simeq 3(1+z)R_{\rm tr} = 0.8~{\rm pc}\left(\frac{1+z}{1+z_{\rm eq}}\right)^{-\frac{1}{3}}\left(\frac{M_{\rm PBH}}{M_{\oplus}}\right)^{\frac{1}{3}}.
\eeq
 In the relevant region $r \simeq R_{\rm eq}$, a large part of the halo is made up of the WIMPs accreted at $z \simeq z_{\rm eq}$. Therefore, if the PBH mass is taken as,
\beq
	M_{\rm PBH} \lsim M_{\oplus}\left(\frac{m_{\rm WIMP}}{100~{\rm GeV}}\right)^{-\frac{3}{2}}\left(\frac{T_{\rm kd}}{10~{\rm MeV}}\right)^{-\frac{3}{2}},
\eeq
the gravitational velocity dispersion is exponentially suppressed due to the free streaming of the WIMPs and can be safely neglected.

Next, we consider the thermal velocity dispersion. The WIMPs have the thermal velocity dispersion $\sqrt{3T_{\rm kd}/m_{\rm WIMP}}$ at the kinetic decoupling and the velocity varies as $\propto (1+z)$ after that. Hence, the thermal velocity dispersion is given by,
\beq \label{eq:thermal}
	\sigma_{\rm t} \simeq 35~{\rm cm/s}\left(\frac{m_{\rm WIMP}}{100~{\rm GeV}}\right)^{-\frac{1}{2}}\left(\frac{T_{\rm kd}}{10~{\rm MeV}}\right)^{-\frac{1}{2}}\left(\frac{1+z}{1+z_{\rm eq}}\right).
\eeq
 The orbit can be considered as quasi-radial at $r \simeq R_{\rm eq}$ if $R_{\rm eq}$ is larger than the radius of the circular motion $(3R_{\rm tr}\sigma_{\rm t})^2/GM_{\rm PBH}$, hence the PBH mass should satisfy
\beq
	M_{\rm PBH} \gsim 6 \times 10^{-4}M_{\oplus}\left(\frac{m_{\rm WIMP}}{100~{\rm GeV}}\right)^{-\frac{3}{2}}\left(\frac{T_{\rm kd}}{10~{\rm MeV}}\right)^{-\frac{3}{2}}.
\eeq
In conclusion, the angular momentum of the WIMPs can be safely neglected if the PBH mass lies in the range $10^{-4}M_{\oplus}-M_{\oplus}$. Note that the WIMP with the thermal velocity $\sigma_{\rm t}$ has pericenter $(3R_{\rm tr}\sigma_{\rm t})^2/2GM_{\rm PBH}$, hence they are not absorbed by the central PBH if $M_{\rm PBH}<10^9M_{\oplus}(m_{\rm WIMP}/100~{\rm GeV})^{-3/4}(T_{\rm kd}/10~{\rm MeV})^{-3/4}$.

\section{Electron/Positron from WIMP Annihilation in UCMH}\label{sec:boost}
\subsection{Propagation}
The electron/positron propagates through the galaxy obeying the following diffusion equation,
\beq \label{eq:diffusion}
\frac{\partial}{\partial E} \left( b(E) f\right) + \nabla\cdot \left( D(E) \nabla f \right) + Q(\vec{x},E) =0,
\eeq
where $f$ is the $e^{\pm}$ number density per unit energy, $b(E)=b_0 (E/1{\rm GeV})^2$ is the rate of energy loss, 
$D(E)=D_0\left( E/1{\rm GeV} \right)^{\delta}$ is the diffusion coefficient, and $Q(\vec{x},E)$ is the $e^{\pm}$ source term.

 In our setup, the source term has a similar form as in the decaying DM case \cite{Nardi:2008ix};
\beq \label{eq:source}
	Q(\vec{x},E) = n_{\rm UCMH}\Gamma\frac{{\rm d}N_{e^{\pm}}}{{\rm d}E},
\eeq
where ${\rm d}N_{e^{\pm}}/{\rm d}E$ is the $e^{\pm}$ spectrum emitted by a WIMP annihilation and $n_{\rm UCMH}$ is the number density of the UCMHs.

In typical propagation models, propagation length of the electron before losing its most of the energy, $d$, is estimated to be
${\cal O}(1)~{\rm kpc}(E/100~{\rm GeV})^{-\frac{\delta-1}{2}}$. The $e^{\pm}$ flux can be written by the solution of the diffusion equation (\ref{eq:diffusion}) as $F = cf/4\pi$. As in the decaying DM case, the resultant flux is insensitive to the uncertainty in the UCMH profile in the host galactic halo. Only their local abundance is relevant to estimate the $e^{\pm}$ flux.

It can occur in the present circumstances that there are only a few or even no UCMHs within the distance $d$ from the Solar system unlike in the decaying DM case. In this case, the signals are rather similar to those from nearby clumps \cite{Hooper:2008kv}. To check whether this is the case, we estimate here the number of the UCMHs within the $e^{\pm}$ propagation length, $N_{r<d}$. Assuming that the UCMHs distribute proportional to the DM abundance in host the galactic halo as $M_{\rm h}n_{\rm UCMH} = (\Omega_{\rm UCMH}/\Omega_{\rm c})\rho_{\odot}$, where $\Omega_{\rm UCMH}$ is the cosmic abundance of the UCMHs, the result is
\beq
	N_{r<d} \simeq 5 \times 10^{11}\Omega_{\rm UCMH}\left(\frac{1+z_s}{30}\right)\left(\frac{M_{\rm PBH}}{M_{\oplus}}\right)^{-1}\left(\frac{d}{1~{\rm kpc}}\right)^3,
\eeq
which is sufficiently large in the relevant mass range $10^{-4}M_{\oplus}-M_{\oplus}$. Note that the UCMHs have an annihilation rate just necessary to explain the PAMELA anomaly in the nearby clump case, $\Gamma = {\cal O}(10^{36-37})~{\rm s^{-1}}$ \cite{Hooper:2008kv}, when $M_{\rm PBH} \sim 10^{3-4} M_{\odot}$ and there exists naturally one UCMH within $1~{\rm kpc}$ if $\Omega_{\rm UCMH} \sim 2 \times 10^{-(3-2)}(1+z_s)^{-1}$. However, the abundance at the mass $M_{\rm PBH} \sim 10^{3-4} M_{\odot}$ is tightly constrained by the CMB observations as $\Omega_{\rm UCMH} \lsim 10^{-9}$ \cite{Ricotti:2007au}. Moreover, the UCMHs can be formed without producing PBHs in this mass range and they would make the situation worse. Therefore, this low-dense case is improbable.

In the next subsection, we estimate the UCMH abundance necessary to explain the PAMELA anomaly assuming that a large number of the UCMHs are contained within the distance $d$ from the Solar system.

\subsection{Positron Excesses induced by the UCMHs}
The analysis can be carried out in just the same way as in the decaying DM case \cite{Nardi:2008ix}, so that we do not repeat the detailed analysis. Here, we assume that the UCMHs have the monochromatic mass function for simplicity.

Since there are a large number of the UCMHs within the $e^{\pm}$ propagation length, we can consider a smooth distribution of the UCMHs. Assuming the UCMH abundance proportional to the DM abundance as before, the source term (\ref{eq:source}) within the $e^{\pm}$ propagation length is estimated to be
\beq\label{eq:source_est}
	\begin{split}
Q_{\odot,{\rm PBH}} \simeq &{\cal O}(10^{-25})~{\rm cm^{-3}s^{-1}} \times \\
&\quad \Omega_{\rm UCMH}\left(\frac{1+z_s}{30}\right)\left(\frac{m_{\rm WIMP}}{100~{\rm GeV}}\right)^{-2}\left(\frac{\left<\sigma v\right>}{3\times 10^{-26}~{\rm cm^3s^{-1}}}\right)\frac{{\rm d}N_{e^{\pm}}}{{\rm d}E}.
	\end{split}
\eeq
In the case of DM annihilation without enhancement in the UCMHs, on the other hand, the source term is given by
\beq
	\begin{split}
Q_{\odot, \scalebox{3.5}[0.9]{/}\hspace{-1.7em} {\rm PBH}}
 \simeq \quad 1.35 \times 10^{-31}~{\rm cm^{-3}s^{-1}}\left(\frac{m_{\rm WIMP}}{100~{\rm GeV}}\right)^{-2}\left(\frac{\left<\sigma v\right>}{3\times 10^{-26}~{\rm cm^3s^{-1}}}\right)\frac{{\rm d}N_{e^{\pm}}}{{\rm d}E}.
	\end{split}
\eeq
Therefore, the boost factor is estimated to be
\beq\label{eq:bf}
	BF = {\cal O}(10^6) \times \Omega_{\rm UCMH}\left(\frac{1+z_s}{30}\right),
\eeq
irrespective of the value of the PBH mass. The boost factor is determined by the UCMH abundance. In the relevant range of the PBH mass $10^{-4}M_{\oplus}-M_{\oplus}$, the UCMHs have mass in the range $10^{-2}M_{\oplus}-10^{2}M_{\oplus}$. In this mass range, the UCMH abundance is constrained by microlensing experiments \cite{DM, Mack:2006gz, Carr:2009jm} as
\beq
	\Omega_{\rm UCMH} < 
	\begin{cases}
		\Omega_{\rm DM} & (10^{-2}M_{\oplus}< M_{\rm h} <10^{-1}M_{\oplus}), \\
		0.1\Omega_{\rm DM} & (10^{-1}M_{\oplus}< M_{\rm h} <10^{2}M_{\oplus}).
	\end{cases}
\eeq
Therefore, the boost factor can be as large as ${\cal O}(10^{4-5})$ in this scenario\footnote{In the case that the accretion stops before $z=30$, we can obtain a larger boost factor, ${\cal O}(10^{6-7})$, as indicated by a factor $(1+z_s)/30$ in Eq.(\ref{eq:bf}). This is because the UCMH mass is smaller by a factor $30/(1+z_s)$ in this case and larger number density of the UCMHs is required to explain their present energy density.}.

 The signals from the source (\ref{eq:source_est}) is similar to that from the decaying DM with mass $2m_{\rm WIMP}$. To compare with the results obtained for the decaying DM, it would be convenient to introduce a corresponding decay rate, $\Gamma_c$, defined as,
\beq
	\Gamma_c \equiv \frac{n_{\rm UCMH}}{\rho_{\odot}/2m_{\rm WIMP}}\Gamma,
\eeq
which is estimated to be
\beq
	\Gamma_c = {\cal O}(10^{-22})~{\rm s^{-1}} \times \Omega_{\rm UCMH}\left(\frac{1+z_s}{30}\right)\left(\frac{m_{\rm WIMP}}{100~{\rm GeV}}\right)^{-1}\left(\frac{\left<\sigma v\right>}{3\times 10^{-26}~{\rm cm^3s^{-1}}}\right),
\eeq
or in terms of the DM life-time, $\tau_c \equiv \Gamma_c^{-1}$,
\beq
	\tau_c = {\cal O}(10^{22})~{\rm s} \times \Omega_{\rm UCMH}^{-1}\left(\frac{1+z_s}{30}\right)^{-1}\left(\frac{m_{\rm WIMP}}{100~{\rm GeV}}\right)\left(\frac{\left<\sigma v\right>}{3\times 10^{-26}~{\rm cm^3s^{-1}}}\right)^{-1}.
\eeq
 As seen in Ref.~\cite{Nardi:2008ix,Chen:2009uq}, the decaying DM with the life-time, $\tau \sim 10^{27}~ {\rm s}(m_{\rm WIMP}/100~{\rm GeV})^{-1}$, can explain the PAMELA anomaly in the WIMP mass range $10^{2-4}~{\rm GeV}$ without violating gamma-ray constraints, though the details depend on the DM decay modes and the galactic halo profile. From these, we can conclude that the UCMH abundance
\beq \label{eq:abundancez}
	\Omega_{\rm UCMH} = {\cal O}(10^{-5})\left(\frac{1+z_s}{30}\right)^{-1}\left(\frac{m_{\rm WIMP}}{100~{\rm GeV}}\right)^2\left(\frac{\left<\sigma v\right>}{3\times 10^{-26}~{\rm cm^3s^{-1}}}\right)^{-1},
\eeq
is necessary to explain the PAMELA anomaly. Note that the high-dense condition, $N_{r<d} \gg 1$, is satisfied for these values of the UCMH abundance.

For the WIMP mass $m_{\rm WIMP} \lsim 10^4~{\rm GeV}$, Eq.(\ref{eq:abundancez}) indicates that a small amount of the UCMHs is sufficient to explain the PAMELA anomaly\footnote{Note that $(1+z_s)/30$ is always larger than unity.}. Hence, Eq.(\ref{eq:abundancez}) can be written in a simpler form as
\beq \label{eq:abundance}
	\Omega_{\rm UCMH} = {\cal O}(10^{-5})\left(\frac{m_{\rm WIMP}}{100~{\rm GeV}}\right)^2\left(\frac{\left<\sigma v\right>}{3\times 10^{-26}~{\rm cm^3s^{-1}}}\right)^{-1},
\eeq
which is much lower than the bounds on the UCMH/PBH abundance \cite{Mack:2006gz, Carr:2009jm} for the most part of the WIMP mass range.

On the other hand, the UCMHs have to constitute a large fraction of the DM, $\Omega_{\rm UCMH} \simeq \Omega_{\rm c}$, for the WIMP mass $m_{\rm WIMP} \sim 10^4~{\rm GeV}$. In this case, Eq.(\ref{eq:abundancez}) reduces to a condition on the redshift when the accretion stops, or the initial fraction of the PBHs in the DM energy density:
\beq \label{eq:zstop}
	\frac{f_{{\rm PBH},i}}{1-f_{{\rm PBH},i}} = {\cal O}(10^{-2})\left(\frac{m_{\rm WIMP}}{10^4~{\rm GeV}}\right)^2\left(\frac{\left<\sigma v\right>}{3\times 10^{-26}~{\rm cm^3s^{-1}}}\right)^{-1}.
\eeq	
In this case, the UCMH mass should be in the range $10^{-8}M_{\oplus}-10^{-2}M_{\oplus}$, where compact objects like the UCMHs  are still allowed to be the DM \cite{Mack:2006gz, Carr:2009jm}. For the WIMP mass $m_{\rm WIMP} \sim 10^4~{\rm GeV}$, this UCMH mass range corresponds to the PBH mass range $10^{-10}M_{\oplus}-10^{-4}M_{\oplus}$. In a lower part of the mass range, we should consider the angular momentum of the WIMPs to obtain a more accurate result.

\section{Prospects and Constraints: Gamma-ray point source}\label{sec:point}

We enclose our paper with a comment on a possibility to observe or have observed the UCMHs as gamma ray point sources; the UCMHs are so bright and sparsely distributed that they could be observed as gamma ray point sources.

\begin{figure}[h!]
	\centering
	\includegraphics[width=.9\linewidth]{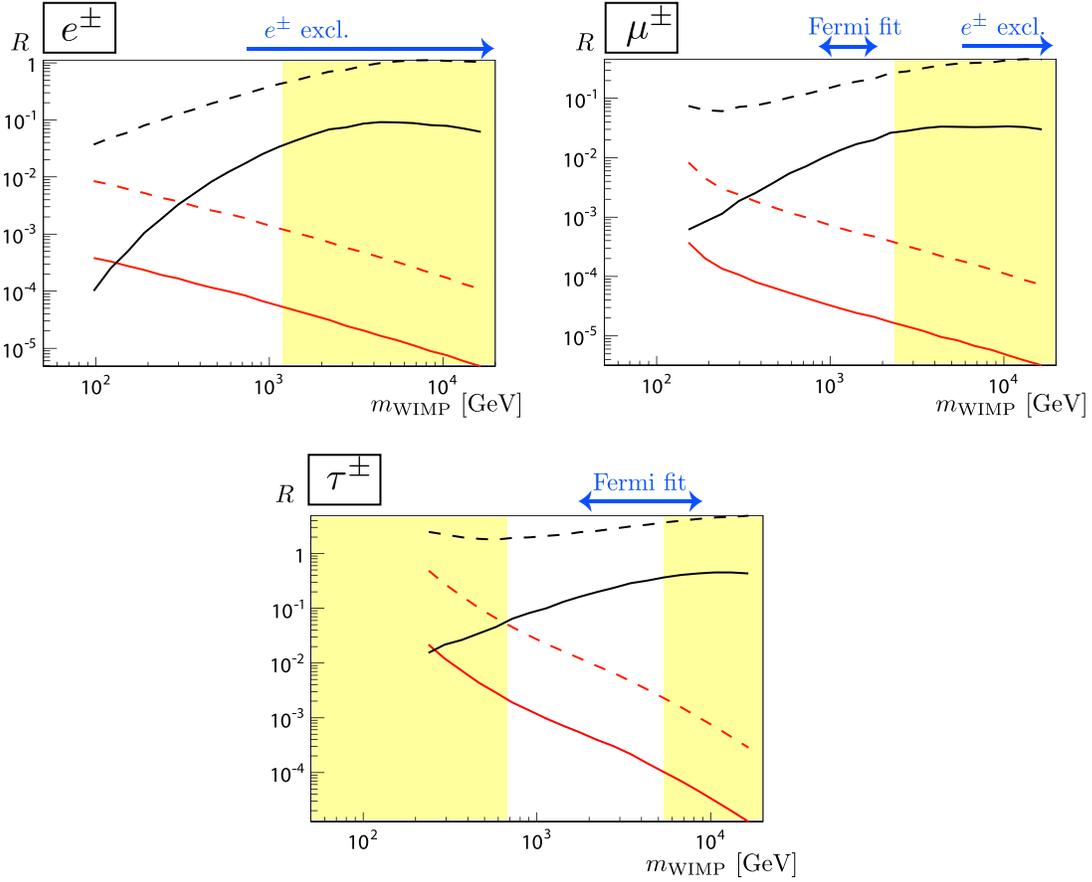}
	\caption{The expected number of the UCMHs observed as point sources by present/future gamma-ray experiments \cite{Buckley:2008ud} for leptonic annihilation modes. Solid lines indicate the expected number observed by HESS (black) and Fermi (red), and dashed lines the number expected for ground-based (black) and space (red) gamma-ray experiments in the future. Yellow shaded regions are excluded by the Fermi observations of the isotropic diffuse gamma-ray \cite{Abdo:2010nz}. ``$e^{\pm}~\text{excl.}$'' shows the region in which $e^{\pm}$ total flux exceeds the Fermi observations \cite{collaboration:2010ij}. The PBH mass is fixed to be the earth mass, $M_{\oplus}$. The lower value of the PBH mass results in the lower values of $R$ as indicated by Eq.(\ref{eq:R}).}
	\label{fig:R}
\end{figure}

The flux of the gamma ray from the UCMH is given by $\Phi_{\gamma}=k_{\gamma}\Gamma/4\pi r^2$, 
where $r$ is distance to the point source. We have also denoted the expected number of photons per an annihilation process as $k_\gamma$, which has values in the range $0.1-10^2$ depending on annihilation mode and energy range of the photons. The UCMH can be observed as a point source if it lies within $d_{\rm point} \equiv (k_{\gamma}\Gamma/4\pi\Phi_{\rm obs})^{1/2}$, where $\Phi_{\rm obs}$ is the point source sensitivity of gamma-ray observations. Then, expected number of the UCMHs observed as point sources is estimated as $R = (4\pi d_{\rm point}^3/3)n_{\rm UCMH}$. Assuming the annihilation rate, $\Gamma$, to be large enough to explain the PAMELA anomaly, we can estimate this number as
\beq \label{eq:R}
	R = {\cal O}(10^{-5})\left(\frac{k_{\gamma}^{-1}\Phi_{\rm obs}}{10^{-8}~{\rm cm^{-2}s^{-1}}}\right)^{-\frac{3}{2}}\left(\frac{m_{\rm WIMP}}{100~{\rm GeV}}\right)^{-1}
\left(\frac{M_{\rm PBH}}{M_{\oplus}}\right)^{\frac{1}{2}}
\left(\frac{\left<\sigma v\right>}{3\times 10^{-26}~{\rm cm^3s^{-1}}}\right)^{\frac{1}{2}}.
\eeq
Typical values of the sensitivity, $\Phi_{\rm obs}$, is $\text{a few} \times 10^{-8}{\rm cm^{-2}s^{-1}}$ for EGRET \cite{Hartman:1999fc} and $\text{a few} \times 10^{-9}{\rm cm^{-2}s^{-1}}$ for Fermi \cite{FERMI}, respectively. Therefore, it is concluded that the UCMHs are unlikely to be observed as point sources by the present experiments. In Fig.\ref{fig:R}, we have displayed the values of $R$ for leptonic annihilation modes obtained by using differential sensitivities of present/future gamma-ray experiments \cite{Buckley:2008ud}. Though the UCMHs is unlikely to be observed by the present experiments as noted above, it will be possible by experiments in the future depending on the nature of the WIMPs and the PBH mass.

\section{Summary and Discussion}\label{sec:summary}
In this paper, we have discussed that a small amount of the PBHs with sub-earth mass can account for the large boost factor to explain the PAMELA anomaly. The boost factor is determined by the present PBH abundance and can be as large as ${\cal O}(10^5)$  in our scenario if the present PBH abundance is comparable to that of the DM. If the large boost factor is originated from particle physics such as the Sommerfeld enhancement, there are severe constraints from cosmology. In addition, the recent Fermi observation of the extra galactic gamma ray gives strong constraints on such a DM \cite{Abdo:2010dk}. In contrast, the signal is very similar to that from the decaying DM in our scenario, which explain the PAMELA anomaly without violating such constraints \cite{Chen:2009uq}. We have also discussed prospects and constraints from the gamma-ray point source experiments and concluded that the future gamma-ray experiments can detect the UCMHs as point sources depending on the nature of the WIMPs and the PBH mass.

In our scenario, the cosmic ray signal from the DM is determined by WIMP nature, such as the mass, annihilation cross section and mode, and also the PBH abundance. The nature of the WIMP can be clarified by the up-coming collider experiment LHC and ILC \cite{Nojiri:2005ph,Baltz:2006fm}. Moreover, the PBH abundance can be also determined by observation of the gravitational wave \cite{Saito:2008jc}.
If there exist a fraction of the PBHs with mass $10^{-4}M_{\oplus}-M_{\rm \oplus}$ as required in our scenario, the gravitational waves associated with the PBHs can be detected by the planned space-based interferometers, such as LISA \cite{LISA}, BBO \cite{BBO} and DECIGO \cite{DECIGO}. Our scenario is testable by the collider and astrophysical experiments.

\section*{Acknowledgements}
We are grateful to T.\ Yanagida and J.\ Yokoyama for careful reading the paper and useful comments
. R.S also thanks A.\ Taruya for useful discussions. This work is supported in part by JSPS
Research Fellowships for Young Scientists and by
World Premier International Research Center Initiative, MEXT, Japan.

\end{document}